# Coverless Information Hiding Based on Generative adversarial networks

Ming-ming LIU    Min-qing ZHANG    Jia LIU    Ying-nan ZHANG    Yan KE

*Abstract*—Traditional image steganography modifies the content of the image more or less, it is hard to resist the detection of image steganalysis tools. To address this problem, a novel method named generative coverless information hiding method based on generative adversarial networks is proposed in this paper. The main idea of the method is that the class label of generative adversarial networks is replaced with the secret information as a driver to generate hidden image directly, and then extract the secret information from the hidden image through the discriminator. It's the first time that the coverless information hiding is achieved by generative adversarial networks. Compared with the traditional image steganography, this method does not modify the content of the original image. therefore, this method can resist image steganalysis tools effectively. In terms of steganographic capacity, anti-steganalysis, safety and reliability, the experimen shows that this hidden algorithm performs well
*Index Terms*—Information security, Coverless Information hiding, Generative adversarial networks.

## 1 Introduction

In the way of traditional steganography, the secret information is embedded into the carriers which is used as the cover of secret information by an invisible way, so as to achieve the purpose of secret communication. The carriers include common digital images, audios, videos and so on [1]. Mostly, traditional steganography directly modifies the carrier to embed secret data. Owing to the partial distortion of the carrier, the third party can detect the existence of hidden secret information by finding the statistic evidence introduced by the embedding method.

 Among all the carriers of information hiding, digital images are used most widely. In traditional image steganography, pixel values are modified to achieve the embedding of secret information. According to the different ways of hiding, the common steganography methods can be classified into two categories: hiding methods in spatial domain and transform domain. Steganography in the spatial domain, such as the method proposed in paper[2] which replace the LSB (least significant bits) of the image with secret data, the adaptive LSB hiding method[3], the spatial adaptive steganography algorithm S-UNIWARD [4],HUGO[5],WOW[6] and so on; The transform domain method is to modify the host image data to change some statistical features to achieve data hiding, such as the hidden method in DFT(discrete Fourier transform) domain [7], DCT (discrete cosine transform) domain [8], and DWT (discrete wavelet transform) domain [9].

(Ming-ming Liu, e-mail:solomon-ming@foxmail.com).

These methods modify the carrier images to embed the secret information by the certain rules, it is inevitable to leave some traces of modification on the carrier. Hence, we are facing such a problem: these steganography methods cannot resist the detection of existing steganalysis tools. In order to resist the detection of all kinds of steganalysis algorithms fundamentally, researchers have proposed the concept of coverless information hiding[10].

Compareing with traditional information hiding methods, Coverless information hiding does not require extra carriers , but generates or obtains the digital images driven by the secret data directly. Zhou Zhili et al [10] proposed a coverless information hiding based on Bag-of-Words model of image, his idea is to establish the mapping relationship between the original image and the secret information. But This method only avoids the change of embedded information to the image carrier, and the capacities of this method are small while the image database is very large. To solve this problem, we propose a novel method----generative coverless information hiding method----which is based on Generative adversarial networks in this paper.

In the recent research on Generative Adversarial Networks (GANs) [11] the generation of image samples is driven by noise directly, which is in good agreement with the idea of coverless information hiding. Therefore, we propose a new generative coverless information hiding method: First, the text is encoded and then as a driver, it combined with the encoded information and the noise to generate image samples. The generated image samples are the hidden image for transmissionin order to realize the generative coverless information hiding. Because the carrier image is not modified, the third party will not easily perceive anomalies about stego images. Furthermore, this method can resist detection of all the existing steganalysis tools.

## 2 Generative adversarial networks

GAN is a very efficient generation model proposed by Goodfellow et al [11] in 2014. The structure of GAN is shown in Figure 1. The idea of GAN comes from the two-person zero-sum game in game theory, whose main structure consists of a generator  and a discriminator, any differentiable function can be used to represent GAN generator (G) and discriminator (D) [12].

Generative Adversarial Networks (GAN) is a recent approach to deep unsupervised learning, proposed in 2014 in [11], which is capable of dynamically representing a sampler from input data distribution and generate new data samples.The main idea of such approach to learning is that two neural networks are trained simultaneously(suppose GAN's inputs are real data X and Random variable Z):

- a generative model (G) that receives noise from the prior distribution pnoise(z) on input and
  transforms it into a data sample from the distribution pg(x) that approximates pdata(x);
- a discriminative model (D) which tries to detect if an object is real or generated by G,if the input of the discriminator is from real data, labelit with 1; if the input sample is from *G(z)*, label it with 0.

The learning process can be described as a minimax game: the discriminator D maximizes the expected log-likelihood of correctly distinguishing real samples from fake ones, while the generator G maximizes the expected error of the discriminator by trying to synthesize better images. Therefore during the training GAN solve the following optimization problem:

$$L(D,G) = E_{x \sim p_{data}(x)}[\log D(x)] + E_{z \sim p_{noise}(x)}[\log(1 - D(G(z)))] \to \min_G \max_D \quad (1)$$

where D(x) represents the probability that x is a real image rather then synthetic, and G(z) is a synthetic image for input noise z.

Coupled optimization problem (1) is solved by alternating the maximization and minimization steps: on each iteration of the mini-batch stochastic gradient optimization we first make a gradient ascent step on D and then a gradient descent step on G. If by θM we denote the parameters of the neural network M, then the update rules are:

- Keeping the G fixed, update the model D by $\theta_D \leftarrow \theta_D + \gamma_D \nabla_D L$ with

$$\nabla_D L = \frac{\partial}{\partial \theta_D} \{E_{x \sim p_{data}(x)}[\log D(x, \theta_D)] + E_{z \sim p_{noise}(z)}[\log(1 - D(G(z, \theta_G), \theta_D))]\} \quad (2)$$

- Keeping D fixed, update G by $\theta_G \leftarrow \theta_G - \gamma_G \nabla_G L$ where

$$\nabla_G L = \frac{\partial}{\partial \theta_G} E_{z \sim p_{noise}(z)}[\log(1 - D(G(z, \theta_G), \theta_D))] \quad (3)$$

The original GAN model has some problems such as unconstrained, uncontrollable, difficult to interpret the noise signal Z. In recent years, many variant GAN models have been derived based on the original GAN model:

- DCGAN [13]: Document [13] extends GAN's idea to DCGAN, which is devoted to image generation. In this paper, the advantages of confrontation training in image recognition and generation are discussed and put forward Methods to build and train DCGANs. In [RMC15] the GAN idea was extended to deep convolutional networks (DCGAN), which are specialized for image generation. The paper discusses the advantages of adversarial training in image recognition and generation, and give recommendations on constructing and training DCGANs.
- Conditional GAN [14]: Can generate the target of the specified category.
- InfoGAN [15]: One of the five breakthroughs that OPENAI calls of 2016, which achieves an efficient use of noise Z and correlates the specific dimension of Z with the semantic features of the data to derive an interpretable characterization .

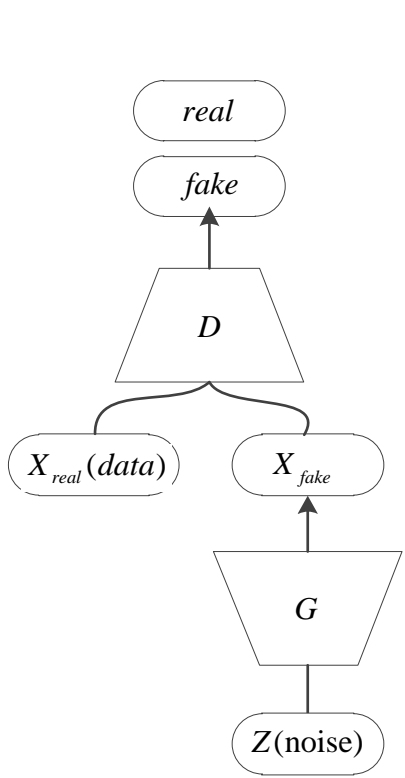 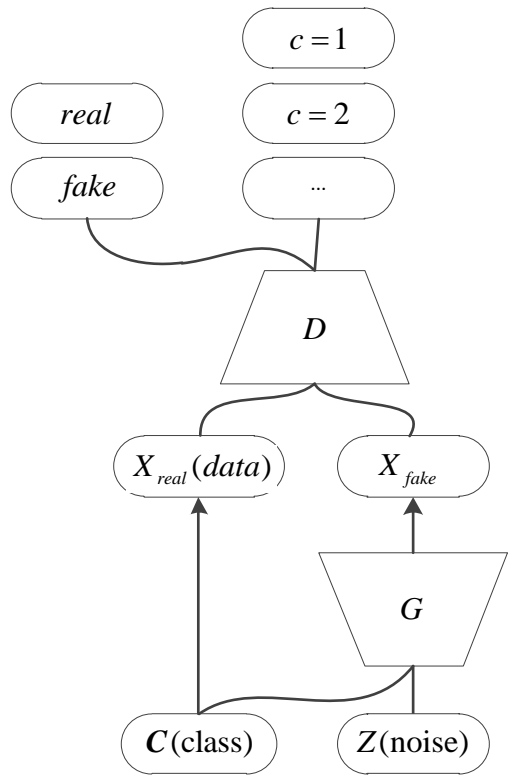

Figure 1: The structure of GAN    Figure 2: The structure of ACGAN

## 3 Carrierless Information Hiding Based on ACGAN

### 3.1 ACGAN

ACGAN（auxiliary classifier GAN）is also a variant of GAN and its structure is shown in Figure 2. In [16], it is proposed that category tags be input to both generators and discriminators on the basis of GAN, thereby not only generating images of a specified category when generating image samples , While tags of this category can also help the discriminator to extend the loss function and enhance the performance of the entire network against attacks.

An additional auxiliary decoder network is added to the AC-GAN discriminator to output the probability of the corresponding category label, and then the loss function is modified to increase the probability of correctly predicting the category.Furthermore, this auxiliary decoding can use the pre-trained discriminators (such as image classifiers) to improve the image steganalysis performance.

Every generated sample in ACGAN has a corresponding class label, $c \sim p_c$ in addition to the noise z. G uses both to generate images $X_{fake} = G(c, z)$. The discriminator gives both a probability distribution over sources and a probability distribution over the class labels, $P(S | X); P(C | X) = D(X)$. The objective function has two parts: the loglikelihood of the correct source, $L_S$, and the log-likelihood of the correct class, $L_C$.

$$L_S = \mathrm{E}[\log P(S = real \mid X_{real})] + \mathrm{E}[\log P(S = fake \mid X_{fake})] \quad (4)$$

$$L_C = \mathrm{E}[\log P(C = c \mid X_{real})] + \mathrm{E}[\log P(C = c \mid X_{fake})] \quad (5)$$

D is trained to maximize LS + LC while G is trained to maximize LC - LS. AC-GANs learn a representation for z that is independent of class label [11]

The structure of ACGAN is similar to the structure of an existing GAN but the training process is more stable than the standard GAN. And the model improves the retention during early training Conditions whenincreasing the number of categories will reduce the quality of the output of the model. The model can generate more categories of designated images, which provides the conditions for the realization of coverless information hiding in this paper.

**3.2 The method of this paper**

Considering that the ACGAN's generator can use the noise Z and the class label C as a driver to generate image samples, and the class label C can be a plurality of classes, In combination with the coverless information hiding, the secret information is directly driven to generate a hidden carrier , A coverless information hidden method based on ACGAN is proposed. The class label C is replaced by the text information K, and the encrypted image is generatedby the text information K to realize the coverless information hiding.

The ACGAN-based coverles information hiding method proposed in this paper consists of the following parts:
1) code dictionary---- the mapping of Chinese characters and category labels library. The function of constructing the dictionary is to convert the text information to be hidden into the corresponding category label sequence, so that both parties can use the same dictionary to convert the text information and the category label combination reversibly.
2) Information hiding and extraction algorithm
   Before the communication, the sender and receiver agree in advance to train ACGAN using the same random variable Z, the same real sample dataset X, the same category label C and the same training steps to get the same generator and discriminator. These information is kept strictly confidential by both parties.

When hiding, firstly, the hidden text information is divided into words according to the word or word existing in the code dictionary, then 15 words or words are continuously selected to form a set of text information pieces, and then the information is encoded according to the code dictionary into the secret information fragment.Finally enter the secret information fragment into the trained ACGAN, and generate the encryptedimage through the generator for transmission.

At the time of extraction, the received encrypted image is input into a discriminator to output a secret information piece, and the secret information piece is decoded into a corresponding text information piece according to the constructed dictionary, All pieces of text information are connected in sequence to obtain the hidden text information in the received encrypted image.

### 3.2.1 Code dictionary construction

For this algorithm, taking into account the complexity of the calculation, the dictionary should be constructed to cover all the commonly used Chinese characters (that is, 3755 Chinese characters at the national level), as well as the national secondary characters and some commonly used phrases and punctuation to improve the capacity of information hiding. Based on mnist handwritten digit set has 0 to 9 a total of 10 categories of tags, the algorithm selected 10000 categories of tag combinations to build the code dictionary, that is, every 4 numbers ( Can be repeatedly selected) as a group, a total of 10000 groups, each corresponding to a Chinese word or phrase, to build a commonly used Chinese characters (or phrases) and category labels combination of one-to-one corresponding dictionary. At the same time we should be regularly replace the dictionary to reduce using the same dictionary frequency,and increase the difficulty of deciphering.

### 3.2.2 hidden algorithm

Hiding and extracting text information is the key point of information hiding algorithm, as shown in Figure 3, the specific hiding scheme is as follows:

- Step (1): For text information T that needs to be hidden, according to the dictionary, every 15 Chinese characters or phrases are grouped together, and a serial number mark is added to each group of head, then the text information T is divided into n pieces of text information ,which is $T = \{T_1, T_2, \cdots, T_n\}$

- Step (2): According to the constructed dictionary, each piece of text information is encoded into 64 corresponding category labels by looking up the dictionary to form a new piece of secret information, denoted as K

- Step (3): The category label C in the generator is directly replaced by the secret information K, K is input to the pre-trained ACGAN, the generator has trained the weight value, and the generator is combined by K and Z Input, after a series of deconvolution, regularization and other operations generated encrypted image $G(K, z)$ transmission.

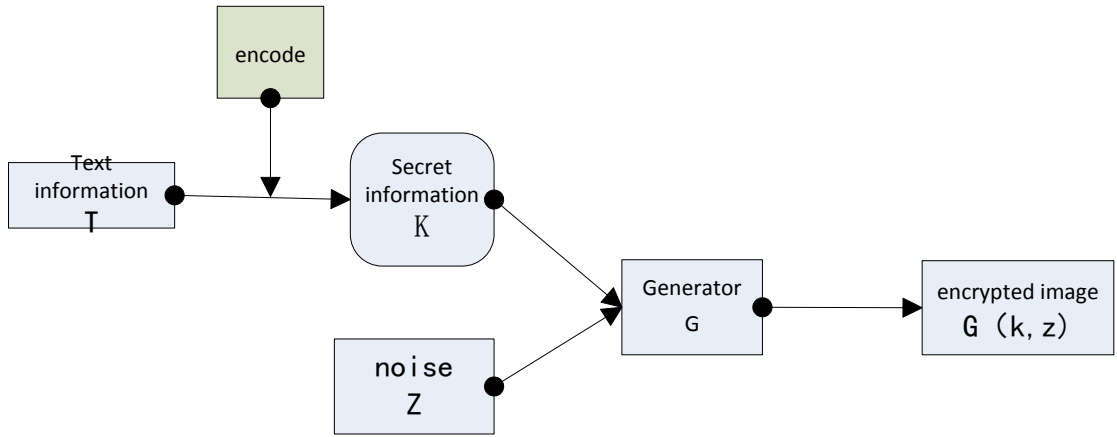

Figure 3: The structure of the hidden algorithm

### 3.2.3 Extraction algorithm

As shown in Figure 4, the algorithm in this paper is as follows:

- Step (1): After the receiver receives the encrypted image $G(K,z)$, it inputs the image $G(K,z)$ into a pre-trained discriminator, and after such as convolution and regularization, the classifier outputs the image logarithm logits.
- Step (2): Use the softmax function to convert the log likelihoods of image categories into the probabilities that the images belong to each category.
- Step (3): Use the argmax function to output the category with the highest probability and extract the category label
- Step (4): Due to network delay and other intentional or unintentional scrambling attacks, the order of the images received by the receiver may be different from the sequence of the images of the hidden text information fragments of the sender. Therefore, firstly, the image corresponding to the received image Secret information K head serial number mark.
- step (5): the secret information K according to the sequence number, according to the constructed dictionary through the table, followed by the secret information K is decoded into the corresponding text message fragments, in order to connect all the text message fragments received To the hidden image with the hidden text information T, in order to achieve coverless information hiding

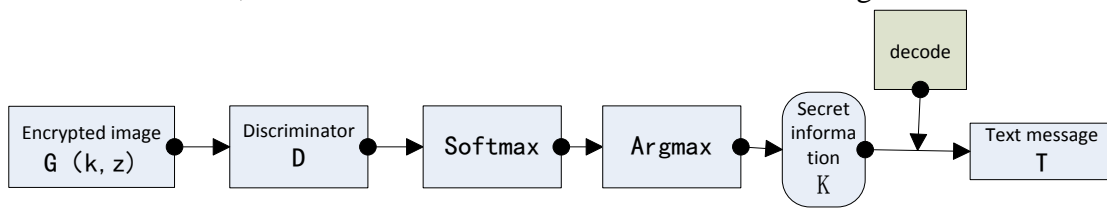

Figure 4: The structure of extraction algorithm

## 4 Algorithm Experiment and Analysis

In this experiment, we assume that the ACGAN network trained by the two parties prior to the communication is as follows: The random noise z is uniformly distributed on (-1,1), and the real sample dataset is a classic handwritten numeral set mnist (containing 60,000 28x28pi Of the handwritten digital grayscale images), the category label for the mnist data set of digital labels 0 to 9, the training steps to 1000. Experimental platform for Google's deep learning platform Tensorflow v0.12, computing graphics for NVIDIA970. To be hidden Text information, randomly selected an article from"Armed Police News,".

By C2D-BN-LR we denote the following structural block of a convolutional neural network: Conv2d---Batch Normalization---Leaky ReLU. The Image Discriminator network have this structure: four C2D-BN-LR layers, then a fully connected layer (1 neuron), Sigmoid function is used to compute an output. The Image generator network G is (in order) a fully-connected layer (8192 neurons), four C2D-BN-LR with Fractional-Strided convolution, then the Hyperbolic tangent function layer is used to compute normalised output.

The SGAN model is trained to solve (4) using the Adam optimization algorithm with the learning rate and update parameters $\beta_1 = 0.5$ and $\beta_2 = 0.999$. For each mini-batch of images we update weights of D and S once, then we update weights of G twice.

### 4.1 The implement of the hidden algorithm and extraction algorithm

In order to facilitate the observation of experimental results in subsequent experiments, it is assumed that the secret information K to be hidden after the encoded text information T is encoded:

$$K = \begin{Bmatrix} 0 & 0 & 0 & 0 & 0 & 0 & 0 & 0 \\ 1 & 1 & 1 & 1 & 1 & 1 & 1 & 1 \\ 2 & 2 & 2 & 2 & 2 & 2 & 2 & 2 \\ 3 & 3 & 3 & 3 & 3 & 3 & 3 & 3 \\ 4 & 4 & 4 & 4 & 4 & 4 & 4 & 4 \\ 5 & 5 & 5 & 5 & 5 & 5 & 5 & 5 \\ 6 & 6 & 6 & 6 & 6 & 6 & 6 & 6 \\ 7 & 7 & 7 & 7 & 7 & 7 & 7 & 7 \end{Bmatrix}$$

Every time we train ACGAN network 10 times in the experiment, we carry out a test that generates secret image by secret message K and extracts secret message K from encrypted image.

The experimental results show that, as shown in Figure 5, after 10 times of ACGAN training, the generated image is nearly white noise, the human eye can not distinguish the number of categories, and the classifier extracted by the discriminator is completely confused. The category labels are as follows: (experimental data is from

all output data):
2017-10-24T10:51:51.528626: step 10, d_loss 5.82928, g_loss 7.95516
('class_loss_fake', array([6, 0, 8, 6, 1, 1, 2, 0, 1, 1, 6, 1, 8, 1, 6, 1, 8, 1, 1, 6, 5, 1, 7, 6, 3, 7, 6, 1, 6, 5, 6, 1, 6, 1, 4, 4, 6, 1, 2, 5, 5, 7, 7, 3, 6, 1, 5, 1, 6, 6, 6, 8, 1, 6, 6, 6, 7, 1, 7, 1, 3, 7, 1, 7]))

  With the increase of training times, the generated images gradually become clear and visible. After 80 trainings, although for the generated image human eyes still find it difficult to distinguish the digital categories, the category labels extracted by the discriminator are basically correct, and the 64 category labels Only 3 errors occurred. After 80 trainings, the category tags extracted are as follows:
2017-10-24T10:52:44.363718: step 80, d_loss 1.55887, g_loss 2.41807
('class_loss_fake', array([0, 0, 0, 0, 0, 0, 0, 0, 1, 8, 1, 1, 1, 1, 1, 1, 2, 2, 2, 2, 2, 2, 2, 2, 3, 3, 3, 3, 3, 3, 3, 3, 4, 4, 4, 4, 4, 4, 4, 4, 5, 5, 5, 5, 5, 5, 5, 8, 6, 6, 6, 2, 6, 6, 6, 7, 7, 7, 7, 7, 7, 7, 7, 7]))

  From the 140th onward, the class labels extracted by the discriminator have been completely no error-, but it is still difficult to generate the images for effective identification of the digital categories. From the 240th onward, the generated images are basically clear and visible, and after the 610th generation The image has the same effect as the 990th training, and the digital category can be clearly identified with the human eye. The categories tagged after 140 times, 240 times, 610 times and 990 times respectively are as follows:
2017-10-24T10:53:35.306073: step 140, d_loss 0.850362, g_loss 2.38015
('class_loss_fake', array([0, 0, 0, 0, 0, 0, 0, 0, 1, 1, 1, 1, 1, 1, 1, 1, 2, 2, 2, 2, 2, 2, 2, 2, 3, 3, 3, 3, 3, 3, 3, 3, 4, 4, 4, 4, 4, 4, 4, 4, 5, 5, 5, 5, 5, 5, 5, 5, 6, 6, 6, 6, 6, 6, 6, 6, 7, 7, 7, 7, 7, 7, 7, 7]))
2017-10-24T10:54:56.809609: step 240, d_loss 1.09391, g_loss 1.56321
('class_loss_fake', array([0, 0, 0, 0, 0, 0, 0, 0, 1, 1, 1, 1, 1, 1, 1, 1, 2, 2, 2, 2, 2, 2, 2, 2, 3, 3, 3, 3, 3, 3, 3, 3, 4, 4, 4, 4, 4, 4, 4, 4, 5, 5, 5, 5, 5, 5, 5, 5, 6, 6, 6, 6, 6, 6, 6, 6, 7, 7, 7, 7, 7, 7, 7, 7]))
2017-10-24T11:00:13.907692: step 610, d_loss 0.899276, g_loss 1.36016
('class_loss_fake', array([0, 0, 0, 0, 0, 0, 0, 0, 1, 1, 1, 1, 1, 1, 1, 1, 2, 2, 2, 2, 2, 2, 2, 2, 3, 3, 3, 3, 3, 3, 3, 3, 4, 4, 4, 4, 4, 4, 4, 4, 5, 5, 5, 5, 5, 5, 5, 5, 6, 6, 6, 6, 6, 6, 6, 6, 7, 7, 7, 7, 7, 7, 7, 7]))
2017-10-24T11:05:50.231032: step 990, d_loss 1.09119, g_loss 0.644074
('class_loss_fake', array([0, 0, 0, 0, 0, 0, 0, 0, 1, 1, 1, 1, 1, 1, 1, 1, 2, 2, 2, 2, 2, 2, 2, 2, 3, 3, 3, 3, 3, 3, 3, 3, 4, 4, 4, 4, 4, 4, 4, 4, 5, 5, 5, 5, 5, 5, 5, 5, 6, 6, 6, 6, 6, 6, 6, 6, 7, 7, 7, 7, 7, 7, 9, 7]))

  Because the goal of this algorithm is to realize the coverless information hiding, the hidden secret information should not be directly recognized by the human eye. Therefore, it is suitable to train 140 times to 240 times, the generate image couldn't Identify the number of categories, but also to ensure that the discriminator to extract the correct category label.

  At the same time, we also found in the experiment that after the 140th time, although most of the extracted category labels can be completely correct, there is a very rare case where a category label will be extracted incorrectly. As shown in FIG. 5, In the training of the 990th times, the discriminator to the category of the last two tags in the error. At this time only need to add the error correction code during encoding can ensure the correctness of the decoding. It can also be seen that the traditional Information hiding is to modify the carrier, does not change the secret information, and in the coverless information hiding, the secret information is allowed to exist in the process of covert communication error, simply by adding error correction code and other ways to ensure the correctly decoding .And because there is no need to make any changes to the carrier, it increases the anti-testability of information hiding. In view of the simplistic mnist handwritten data set, when the generated image become clear, the Category labels can be directly recognized by the human eye, the next step to consider this method applied to celeb A face data sets and other complex natural image set to solve this problem.

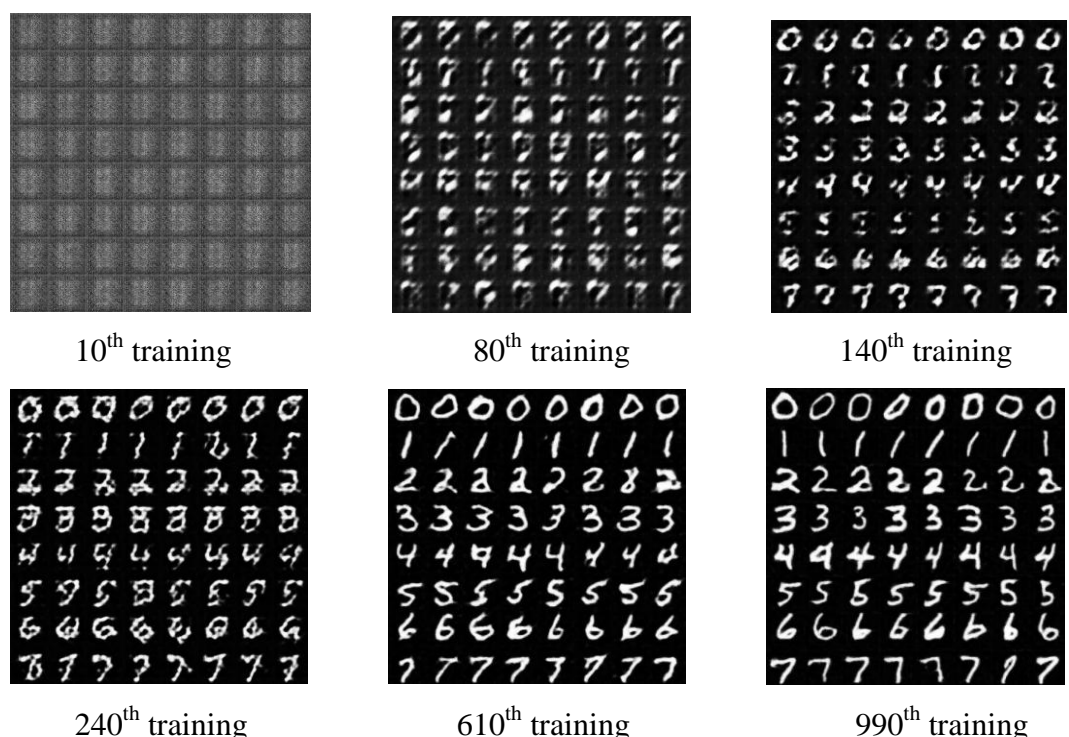

| 10$^{th}$ training | 80$^{th}$ training | 140$^{th}$ training |
| 240$^{th}$ training | 610$^{th}$ training | 990$^{th}$ training |

Figure 5: the generate picture with sicrit

The following experiment and analysis will be the algorithm of each sub-image information hiding capacity, image distortion and anti-test, the reliability of covert communication, the safety with confidential information after exposure.

**4.2 capacity**

Three different word segmentation methods are used to test the information hiding capacity of each sub-image in this paper. The experimental results are shown in Table 1.Do not use the word dictionary in the code table, the text information to be hidden directly into the single word. When hidden, a word corresponds to a category label combination (containing four category labels.) . Hidden text information experiment, Each sub-image has a hidden capacity of 64/4 = 16 words, and since the sub-image header's category label combination is a serial number mark, the information hiding capacity per sub-image is 15 words. without Words dictionary, the number of words in the dictionary and the average word length are 0.

Select the first 100 selected articles from the length of 2 word segmentation dictionary (the number of words is 100, the average word length of 2). At this point, there are both the first and second level national standard word in the Corresponding category label combination and words segmentation dictionary corresponding to the category label combination in the dictionary. according to the positive maximum matching code dictionary principle, that is, if the secret information contains the words of code dictionary, then cut Divided into words, otherwise cut into single word, randomly selected 10 text fragments, the experimental results for each sub-image average hidden capacity of 22.6 words.

Select the first 100 phrases with a length of 3 from the article to create a word segmentation dictionary (the number of words is 100, the average word length is 3.) Using the principle of a positive maximum matching code dictionary, randomly select 10 text segments, the experimental results for each The average hidden capacity of sub-images is 31.3 words.

It can be seen from the experiment that increasing the average length of the phrases in the codelist word segmentation dictionary can increase the average information hiding capacity of each sub-image. From the theoretical analysis, the information hiding capacity of a single sub-image is 15 categories of label combinations corresponding Chinese characters, that is, the length of the secret information segment. the average information hiding capacity of the multi-sub-images $\overline{C}$ is the average number of Chinese characters corresponding to each 15 category label combinations, that is, the average length of the segment of the secret information after the word segmentation:

$$\overline{C} = \frac{\sum_{i=1}^{n} C_i}{n}$$

Where: n is the number of secret information fragments, $C_i$ is the length of the i-th secret information fragment. To improve the information hidden capacity of each sub-image:

- According to the commonly used contents of both communication parties, construct a more complete code dictionary and increase the number of phrases in the code word dictionary, so that the classified information fragments contain as many phrases as possible.
- Because the secret image generated by this method has only 8x8 total 64 category labels, the information hiding capacity of each sub-image can be improved by including more category labels in the generated image within the range of a given pixel (such as 512x512pi).

Table 1 Experimental results for testing the hiding capacity

| Average length of words in dictionary (Chinese characters/words) | Words Numbers of lexical dictionaries | The capacity of literature [10] 's algorithm (Chinese characters/sub-images) | The capacity of our paper's algorithm (Chinese characters/sub-images) |
|---|---|---|---|
| 0 | 0 | 1 | 15 |
| 2 | 100 | 1.57 | 22.6 |
| 3 | 100 | 1.86 | 31.3 |

**4.3 image distortion and anti-testability**

Since the encrypted image in this paper is generated directly by ACGAN, no change is made to the encrypted image in the embedding of the textual information and the transmission of the encrypted image, so that the encryptedimage is not distorted and can effectively resist human eyes Detection, but also fundamentally resist the detection of statistics-based information hiding analysis algorithm.

**4.4 reliability**

In this paper, we study the coverless information hiding. The hidden image that conceals the secret information is a pseudo-natural image generated by ACGAN without any modification. Compared with traditional encryption and information hiding methods, the proposed method is more difficult to cause the suspicion of the aggressor and more covert for secret communications.

**4.5 Security**

Assuming that the attacker has suspected that the image being delivered contains secret information, it is also difficult to extract the secret information from the contained image using the discriminator because it does not have the same ACGAN model as the communicating parties. Even if an attacker accidentally extracts secret information, It does not know the combination of the dictionary and the category label, nor can it decode the secret information into the original text information, thereby ensuring the security of concealed communication.

**5 Conclusion**

In this paper, we proposed a new generative coverless steonagaphy method based on generative adversarial networks. We use the ACGAN that combining noise and secret information to generate indistinguishable hidden-image.Mainly, This method consists of two parts: the construction of the code dictionary, the algorithm of hiding and extracting text information. Experimental and theoretical analysis show that this method not only can resist the existing steganalysis algorithm detection effectively, but also its hidden communication has good reliability and security. We propose a basic framework for the future generative coverless information hiding in this paper. Because the carrier image is not modified, the third party will not easily perceive anomalies about stego images. Furthermore, this method can resist the detection of all the existing steganalysis tools. The resulting image is still not clear enough. The next step is to create a more realistic and clear natural image with a more sophisticated Wasserstein GAN (W-GAN) [17].

**references:**

[1] Shen C X, Zhang H G, Feng D G. A survey of information security [J]. Science China, 2007,37(2): 129-150. (in Chinese)
[2]Tirkel A Z, Rankin G A, Schyndel R V. Electronic watermark [C]//Digital Image Computing, Technology and Applications, 1993.
[3] Yang C H, Weng C Y, Wang S J. Adaptive data hiding in edge areas of images with spatial LSB domain systems [J]. IEEE Transactions on Information Forensics & Security, 2008, 3(3):488-497.
[4] Holub V, Fridrich J, Denemark T. Universal distortion function for steganography in an arbitrary domain[J]. Eurasip Journal on Information Security, 2014, 2014(1):1.
[5] Pevný T, Filler T, Bas P. Using High-Dimensional Image Models to Perform Highly Undetectable Steganography[J]. Lecture Notes in Computer Science, 2010, 6387:161-177.
[6] Holub V, Fridrich J. Designing steganographic distortion using directional filters[C]// IEEE International Workshop on Information Forensics and Security. IEEE, 2013:234-239.
[7] Ruanaidh J J K O, Dowling W J, Boland F M. Phase watermarking of digital images [C]// International Conference on Image Processing, 1996: 239-242.
[8] Cox I J, Kilian J, Leighton F T. Secure spread spectrum watermarking for multimedia [J].IEEE Transactions on Image Processing, 2010, 6(12): 1673-87.
[9] Lin W H, Horng S J, Kao T W. An efficient watermarking method based on significant difference of wavelet coefficient quantization [J]. IEEE Transactions on Multimedia, 2008, 10(5):746-757.
[10] Zhou Z L, Cao Y, Sun X M. Coverless information hiding based on bag-of-words model of image[J]. Journal of Applied Sciences, 2016, 34(5):527-536.
[11]Goodfellow I, Pouget-Abadie J, Mirza M, Xu B, WardeFarley D, Ozair S, Courville A, Bengio Y. Generative adversarial nets. In: Proceedings of the 2014 Conference on Advances in Neural Information Processing Systems 27. Montreal, Canada: Curran Associates, Inc., 2014. 2672–2680
[12] Wang K F, Gou C, Duan Y J, et al. Generative Adversarial Networks:The State of the Art and Beyond[J]. Acta Automatica Sinica, 2017, 43(3):321-332.
[13]Radford A, Metz L, Chintala S. Unsupervised Representation Learning with Deep Convolutional Generative Adversarial Networks[J]. Computer Science, 2015.
[14] Mirza M, Osindero S. Conditional Generative Adversarial Nets[J]. Computer

Science, 2014:2672-2680.

[15] Chen X, Duan Y, Houthooft R, et al. InfoGAN: Interpretable Representation Learning by Information Maximizing Generative Adversarial Nets[J]. Computer Science, 2016.

[16] Odena A, Olah C, Shlens J. Conditional Image Synthesis With Auxiliary Classifier GANs[J]. Computer Science, 2016.

[17] Arjovsky M, Chintala S, Bottou L. Wasserstein GAN[J]. Computer Science, 2017.